\documentstyle[12pt]{article}

\begin{document} 
\pagestyle{plain}

\newcommand{\be}{\begin{equation}}
\newcommand{\ee}{\end{equation}}
\newcommand{\ba}{\begin{eqnarray}}
\newcommand{\ea}{\end{eqnarray}}
\newcommand{\vp}{\varphi}
\newcommand{\pr}{\prime}
\newcommand{\psib} {\bar{\psi}}
\def\ddk{[d^3{\rm k}]}
\def\kk{{\rm k}}
\renewcommand{\theequation}{\arabic{section}.\arabic{equation}}
\makeatletter\@addtoreset{equation}{section}\makeatother
\def\baselinestretch{1.0}
\def\ddk{[d^3{\rm k}]}
\def\kk{{\rm k}}

\begin{titlepage}
\rightline{LBL-38585}
\rightline{March 26, 1996} \vskip .1in
 
\begin{center}
{\large
{Anomalous Transverse Distribution of Pions as a signal for the production
of DCC's}}\\
\ \\\ \\
{\large
F. Cooper,$^1$ ,Y. Kluger,$^{2}$\\
E. Mottola,$^1$ \\
\ \\}
$^1$Theoretical Division \\
Los Alamos National Laboratory\\
Los Alamos, New Mexico 87545, USA\\
$^2$ Nuclear Science Division\\
Lawrence Berkeley Laboratory\\
Berkeley, CA 94720, USA
\ \\
\ \\
\end{center}

\begin{abstract}
We give evidence  that the production
of DCC's during a non-equilibrium phase transition can lead to an
anomalous
transverse distribution of secondary pions when compared
to a more conventional boost invariant hydrodynamic flow in local
thermal
equilibrium. Our results pertain to the linear $\sigma$ model,treated in
leading order in large-$N$, in a boost
invariant approximation. We also show that the interpolating number density  
of the field theory calculation plays the role of a  classical relativistic  
phase space number distribution in determining the momentum distribution of  
pions in the center of mass frame.
\end{abstract}

\end{titlepage}
\section{Introduction}

Recently there has been a growing interest in the possibility of
producing disoriented chiral condensates (DCC's) in a high energy
collision  \cite{Anselm,BjorkenIJMP,RajaWil}. This
idea was first proposed to explain CENTAURO events in cosmic ray
emperiments where there was a deficit of neutral pions
\cite{Cosmicrays}. It was proposed that a nonequilibrium chiral phase
transition such as a quench might lead to
regions of DCC \cite{RajaWil}.

In a previous paper \cite{DCC}, we discussed in detail  the time
evolution of a non-equilibrium
chiral phase transition in  the linear $\sigma$
model in the leading order in large-N approximation.  We showed in that
paper that the expansion into the vacuum of the initial energy
distribution
led to rapid cooling. This caused the system, initially in quasi local
thermal
equilibrium to progress from the unbroken chiral symmetry phase to the
broken symmetry phase vacuum.  This expansion is accompanied by
the exponential growth of
low momentum modes for short periods of proper time for a range of initial  
conditions.  This exponential
growth of long wave
length modes is the mechanism for the production of disordered chiral
condensates. Thus the production of DCC's results in an
enhancement of particle production in the low momentum domain. Whether
such an instability occurs depends
on the size of the initial fluctuation from the initial thermal
distribution.   The relevant
momenta for which this exponential growth occurs are the  transverse
momenta
and the momenta  $k_\eta = -Ez+tp$ conjugate to the fluid rapidity
variable
$\eta = \tanh^{-1}(z/t)$. We also found that the distribution of
particles in
these momenta   had more length scales than found in local thermal
equilibrium. When there is local thermal equilibrium, the length scales
are
the mass of the pion and the temperature which is related to the
changing
energy
density, both of which depend on the proper time $\tau$.

In this paper we reexpress our previous results in terms of the
physically measurable transverse distribution of particles in the
collision center of mass frame.
We  show that whenever DCC's are produced there is a noticable
distortion
of the transverse spectrum, namely an enhancement of particles at low
transverse momentum,  when compared to a local equilibrium evolution. We
will consider two cases, one in which there is exponential growth of
low momentum modes due to the effective pion mass going negative
during the expansion,
and one where the initial fluctuations do not lead to this exponential
growth. Both situations will be compared to a purely hydrodynamical
boost invariant calculation based on local thermal equilibrium.
The idea that the production of DCC's might lead to a distortion of
the transverse distribution of pions is also found in the work
of Gavin and collaborators \cite{Gavin}. However, our calculation is more
realistic in that we
exactly solve the field theory in a large $N$ approximation in the boost  
invariant approximation, using the expansion of the plasma as the cooling  
mechanism for producing instabilities.
In determining the actual spectra of secondaries, we  find that the
adiabatic
number operator of our large-N calculation replaces the
relativistic phase space density $g(x,p)$ of classical transport theory
in
determining the distribution of  pariticles in rapidity and transverse
momentum.
This makes it easy to compare our results with a standard hydrodynamical
calculation in the boost invariant approximation which assumes the final
pions
are in local thermal equilibrium in the comoving frame.

  The model we use to discuss the chiral phase transition is the linear
sigma model described by the Lagrangian:
\begin{equation}
L= {1\over 2} \partial\Phi \cdot \partial\Phi - {1\over 4}
\lambda (\Phi \cdot \Phi - v^2)^2 + H\sigma.
\end{equation}
The mesons  form an $O(4)$ vector $\Phi = (\sigma, \pi_i)$
This can be written in an alternative form by introducing the composite
field:
$\chi = \lambda (\Phi \cdot \Phi-v^2)$.
\begin{equation}
L_2 = -{ 1 \over 2} \phi_i (\Box + \chi) \phi_i + {\chi^2 \over 4
\lambda} +
{1 \over 2} \chi v^2 + H \sigma
\end{equation}
The effective action to leading order in large $N$ is given by
\cite{DCC}
\begin{equation}
\Gamma[\Phi,\chi] = \int d^4x[ L_2(\Phi,\chi,H) + { i \over 2}N {\rm
tr~ln}
G_0^{-1}]
\end{equation}
$$
G_0^{-1}(x,y) = i[\Box + \chi(x)] ~\delta^4(x-y)
$$
Varying the action we obtain:
\begin{equation}
[\Box + \chi(x)] \pi_i = 0 ~~~~ [\Box + \chi(x)]\sigma = H,
\end{equation}
where here and in what follows, $\pi_i$,$\sigma$ and $\chi$ refer to  
expectation values.
Varying the action we obtain
\begin{equation}
\chi= - \lambda v^2 + \lambda (\sigma^2 + \pi \cdot \pi) + \lambda
N  G_0 (x,x).
\end{equation}
If we assume boost invariant kinematics \cite{cfs} which result in flat
rapidity
distributions, then the expectation value of the energy  density is only
a function of the proper time.
The natural coordinates for  boost invariant ($v=z/t$) hydrodynamical
flow
are the fluid
proper time $\tau$  and the fluid rapidity
$\eta$ defined as
$$\tau\equiv(t^2-z^2)^{1/2}, \qquad \eta\equiv{1\over 2}
\log({{t-z}\over{t+z}}).
$$
To implement boost invariance we assume that
mean  (expectation) values of the fields     $\Phi$  and $\chi$ are
functions
of $\tau$ only.
We then get the equations:
\begin{eqnarray}
&&\tau^{-1}\partial_\tau\ \tau\partial_\tau\ \Phi_i(\tau)
 +\ \chi(\tau)\ \Phi_i(\tau) =
 H \delta_{i1} \nonumber \\
&&\chi(\tau) =\lambda\bigl(-v^2+\Phi_i^2(\tau)+
N G_0(x,x;\tau,\tau)\bigr),
\end{eqnarray}
To determine the Green's function $G_0(x,y;\tau,\tau^{\prime})$ we introduce
the auxiliary quantum field $\phi(x,\tau)$ which
obeys the sourceless equation:
\begin{equation}
\Bigl(\tau^{-1}\partial_\tau\ \tau\partial_\tau\ -
\tau^{-2}\partial^2_\eta
-\partial^2_\perp + \chi(x)\Bigr)
\phi(x,\tau)=0.
\end{equation}
$$
    G_0 (x,y;\tau, \tau^{\prime}) \equiv <T\{ \phi(x,\tau)  
~\phi(y,\tau^{\prime})\}>.
$$

We expand the quantum fields in an orthonormal basis:
$$
\phi(\eta,x_\perp,\tau)\equiv{1\over{\tau^{1/2}}} \int \ddk\bigl(\exp(ik
x)
f_\kk(\tau)\ a_\kk\ + h.c.\bigr)
$$
where $k x\equiv k_\eta \eta+\vec k_\perp \vec x_\perp$,
$\ddk\equiv dk_\eta d^2k_\perp/(2\pi)^3$.
The mode functions and $\chi$  obey:
\begin{equation}
\ddot f_\kk +
\bigl({k_\eta^2\over{\tau^2}}+\vec k_\perp^2 + \chi(\tau) +
{1\over{4\tau^2}}\bigr)
f_\kk=0.\label{eq:mode}
\end{equation}
\begin{equation}
\chi(\tau) = \lambda\Bigl(-v^2+\Phi_i^2(\tau)+{1\over \tau}N \int \ddk
|f_\kk(\tau)|^2\  (1+2\ n_\kk) \Bigr) \label{eq:chi}.
\end{equation}

We notice that when $\chi$ goes negative, the low momentum modes with
$${k_\eta^2 +1/4 \over{\tau^2}}+\vec k_\perp^2  < | \chi |$$
grow exponentially. However these modes then feed back into the $\chi$
equation
and this exponential growth then gets damped. It is these growing
modes that lead to the possiblity of growing domains of DCC's as well
as a modification of the low momentum distribution of particles from
a thermal one. The parameters of the model are fixed by physical data.
The PCAC condition is $$
\partial_{\mu} A_{\mu}^i (x) \equiv f_{\pi} m_{\pi}^2 \pi^i(x) = H  \pi^i(x).
$$
In the vacuum state $ \chi_0 \sigma_0= m_{\pi}^2 \sigma_0=H$,
so that $\sigma_0=f_{\pi}$= 92.5 MeV.  The vacuum  gap equation is
$$
m_{\pi}^2 = - \lambda v^2 + \lambda f_{\pi}^2 +
 \lambda N \int_0^{\Lambda} \ddk {1\over 2\sqrt{k^2+m_{\pi}^2}}.
$$
This leads to the mass renormalized gap equation:
\begin{eqnarray}
&&\chi(\tau)-m_{\pi}^2 = -\lambda f_{\pi}^2+ \lambda \Phi_i^2(\tau) \nonumber \\
&&+{\lambda \over \tau}N
\int \ddk
\{|f_\kk(\tau)|^2\  (1+2\ n_\kk) - {1\over 2\sqrt{k^2+m^2}} \}.
\end{eqnarray}
$\lambda$ is chosen to fit low energy scattering data as discussed in \cite{DCC}.

If we assume that the inititially (at $\tau_0 = 1$) the system is in local  
thermal equilibrium in a comoving frame we have
\begin{equation}
n_k = {1 \over e^{\beta_0  E^0 _k} -1}
\end{equation}
where $ \beta_0 = 1/T_0$ and $E^0_k=
\sqrt{{k_\eta^2\over{\tau_0^2}}+\vec k_\perp^2 + \chi(\tau_0)}$.

The initial value of
$\chi$ is determined by the equilibrium gap equation for an initial
temperature of $ 200 MeV$ and is $.7 fm^{-2}$ and the initial value of  
$\sigma$ is just ${H \over \chi_0 }$. The phase transition in this model  
occurs at a critical
temperature
of $160 MeV$. To get into the unstable domain, we then introduce fluctuations  
in the time derivative of the classical field.  We
varied the value of the initial proper time derivative of the sigma
field expectation value and found that for $\tau_0 = 1 fm$ there is a narrow  
range of initial
values that
lead to the growth of instabilities, namely
$.25 <  \vert \dot{\sigma} \vert   < 1.3$.

Figure one displays the results of the numerical simulation for the
evolution of the system (\ref{eq:mode})--(\ref{eq:chi}).
We display the auxiliary field $\chi$ in units of $fm^{-2}$
and the proper time
in units of $fm$   ($ 1fm^{-1} = 197
MeV$) for two simulations, one with an instability
($\dot\sigma|_{\tau_0}=-1$)
and one without  ($\dot\sigma|_{\tau_0}=0$).
We notice that for both initial conditions, the system eventually
settles down to the broken symmetry vacuum result as a result of the
expansion.

To determine the spectrum of particles we introduce the
interpolating number density which is defined by expanding the fields in
terms of mode
functions $f^0_k$ which are first order in an adiabatic expansion of the
mode
equation.
\begin{equation}
f^0_k = {e^{-iy_k(\tau)} \over \sqrt{2 \omega_k}}; ~~ dy_k/dt =
\omega_k,
\end{equation}
where $\omega_\kk(\tau)\equiv({k_\eta^2/{\tau^2}}+
\vec k_\perp^2 + \chi(\tau))^{1/2}$.
This leads to the alternative expansion of the fields:
\begin{equation}
\phi(\eta,x_\perp,\tau)\equiv{1\over{\tau^{1/2}}} \int \ddk\bigl(\exp(ik
x)
f^0_\kk(\tau)\ a_\kk(\tau)\ + h.c.\bigr)
\end{equation}

The two sets of creation and annihilation operators are connected by
a Bogoliubov transformation:
\begin{equation}
a_k(\tau) = \alpha(k,\tau) a_k + \beta(k,\tau) a^{\dag}_{-k}.
\end{equation}
$\alpha$ and $\beta$ can be determined from the exact time evolving mode
functions via:
\begin{eqnarray}
\alpha(k,\tau) &= & i (f_k^{0\ast} { \partial f_k \over \partial \tau} -
{\partial f_k^{0 \ast}  \over \partial \tau}f_k) \nonumber \\
\beta(k,\tau) &=& i (f_k^{0} { \partial f_k \over \partial \tau} -
{\partial f_k^{0 }  \over \partial \tau}f_k).
\end{eqnarray}
In terms of the initial distribution of particles $n_0(k)$ and $\beta$
we have:
\begin{eqnarray}
n_k(\tau) &\equiv& f(k_{\eta},k_{\perp},\tau)
= < a^{\dag}_k (\tau) a_k (\tau) >\nonumber\\
&=& n_0(k) + |\beta(k,\tau)|^2 ( 1+2n_0(k)).
\end{eqnarray}
$n_k(\tau)$ is the adiabatic invariant
interpolating phase space number density which becomes the actual
particle
number density when interactions have ceased. When this happens the
distribution of particles is
\begin{equation}
f(k_{\eta}, k_{\perp},\tau) = {d^6 N \over \pi^2 dx_{\perp}^2
dk_{\perp}^2
d\eta dk_{\eta}}.
\label{interp}
\end{equation}
We now need to  relate this quantity to the physical spectra of
particles
measured in the lab. At late $\tau$ our system relaxes to the vacuum and
$\chi$
becomes the square of the physical pion mass $m^2$. We introduce the
outgoing
pion particle rapidity   $y$ and
$m_{\perp}= \sqrt{k_{\perp}^2 + m^2}$
defined by the particle 4-momentum in the center of mass coordinate system
\begin{equation}
k_{\mu} = (m_{\perp} \cosh y, k_{\perp},m_{\perp} \sinh y)
\label{kmu}
\end{equation}
The boost that takes one from the center of mass coordinates to
the comoving frame where the energy momentum tensor is diagonal is given
by  $\tanh \eta= v = z/t$, so that one can define the ``fluid"
4-velocity
in the center of mass frame as
\begin{equation}
u^{\mu} = (\cosh \eta,0,0, \sinh \eta)
\end{equation}
We then find that the variable
\begin{equation}
\omega_k = \sqrt{m_{\perp}^2+{ k_{\eta}^2 \over \tau^2} } \equiv
k^{\mu}u_{\mu}
\end{equation}
 has the meaning of the energy of the particle in the comoving
frame.
The momenta  $k_{\eta}$ that enters into the adiabatic phase space
number
density is one of two canoncial momenta to the variables defined by the
coordinate transformation to fluid light cone variables. Namely the
variables
\begin{eqnarray}
 \tau = (t^2-z^2)^{1/2}  \qquad \,
\eta = \frac{1}{2} \ln \left({{t+z} \over{t-z}} \right) \nonumber
\end{eqnarray}
have as their canonical momenta
\begin{eqnarray}
k_{\tau}= Et/ \tau -k_z z/ \tau \quad \, k_{\eta}  = -Ez + t k_z.
\label{boost_4trans}
\end{eqnarray}
To show this we consider the metric $ ds^2= d\tau^2 - \tau^2 d \eta^2$
and the free Lagrangian
\begin{equation}
L = {m \over 2} g_{\mu \nu} {dx^{\mu} \over ds} {dx^{\nu} \over ds}
\end{equation}
Then we obtain for example
\begin{eqnarray}
k_{\tau} &=& m {d \tau \over ds} = m [({\partial \tau \over \partial
t})_z {d t
\over ds}+ ({\partial \tau \over \partial z})_t {d z \over ds}]
\nonumber \\
&=& {Et -k_z z  \over \tau} = k^{\mu} u_{\mu}
\end{eqnarray}
The interpolating phase-space density  $f$ of
particles depends on $k_{\eta}$, ${\bf k}_{\perp}$, $\tau$,  and
it is  found to be $\eta$-independent and given by eq.
(\ref{interp}).

In order to obtain the physical particle rapidity and transverse
momentum
distribution, we change variables from $(\eta, k_{\eta})$ to
$(z, y)$ at a fixed $\tau$
where $y$ is the particle rapidity defined by (\ref{kmu}).
We have
\begin{equation}
E{d^3 N \over d^3 k} =
{d^3N \over  \pi dy\,dk_{\perp}^2 } =
\int\pi dz~ dx_{\perp}^2 ~ J ~ f(k_{\eta},
k_{\perp},\tau)
 \label{boost_J}
\end{equation}
where the Jacobian $J$ is evaluated at a fixed proper time $\tau$
\begin{eqnarray}
 J  &=& \left| \matrix{
 {\partial k_{\eta}}/{\partial y}&{\partial
k_{\eta}}/{\partial z} \cr
{\partial \eta}/{\partial y}&
{\partial \eta}/{\partial z} \cr}\right|
= \left| \frac {\partial k_{\eta}}{\partial y}\frac{\partial\eta}
{\partial z} \right|\nonumber\\
&=&{ m_{\perp} \cosh(\eta-y) \over
\cosh \eta}={\partial k_{\eta}\over \partial z}|_{\tau} .
\label{boost_Jinv}
\end{eqnarray}
We also have
\begin{eqnarray}
k_{\tau} = m_{\perp} \cosh(\eta-y); \quad
k_{\eta} = -\tau m_{\perp} \sinh(\eta-y)\,.
\label{boost_ptaueta}
\end{eqnarray}
 Calling the
integration over the transverse dimension the effective transverse
size of the colliding ions $A_{\perp}$ we then obtain that:
\begin{equation}
{d^3N \over  \pi dy\,dk_{\perp}^2} = A_{\perp} \int dk_{\eta}
f(k_{\eta}, k_{\perp},\tau)
\end{equation}
This quantity is independent of $y$ which is a
consequence of the assumed boost invariance.

In a hydrodynamical model of heavy ion collisions \cite{cfs}, the final
spectra
of pions is given by a combination of the fluid flow and a local thermal
equilibrium distribution in the comoving frame. One calculates this
spectra
at the critical temperature $T_c(x,t)$ when the energy density goes
below
$\epsilon_c = {1 \over (\hbar/mc)^3}.$
This defines the breakup surface, after which the particles no longer
interact
so that this distibution is frozen at that temperature. For an
ultrarelativistic
gas of pions, this occurs when $T_c\approx m$.
The covariant form for the spectra of particles is given by the Cooper-
Frye formula \cite{c-f}:

\begin{equation}
E {d^3N \over d^3k} = {d^3N \over \pi dk_{\perp}^2 dy} =
\int g(x,k) k^{\mu} d \sigma_{\mu}
\end{equation}
Here $g(x,k)$ is the single particle relativistic phase space
distribution
function.
When there is local thermal equilibrium of pions at a comoving
temperature $T_c(\tau)$ one has
\begin{equation}
g(x,k) = { g_{\pi} \over (2 \pi)^3} \{ {\rm exp} [k^{\mu}u_{\mu}/T_c] -1
\}
^{-1}.
\end{equation}
In the  boost invariant approximation, the temperature is only
a function of the proper time $\tau$ and the break up surface is
a constant $\tau$ surface:
\begin{equation}
d\sigma^{\mu} = A_{\perp} (dz,0,0,dt)
= A_{\perp}d\eta (\cosh \eta ,0,0,\sinh \eta )
\end{equation}
where $ A_{\perp} $ is the effective transverse size of the system
following
the heavy ion collision.  Therefore we obtain the relationship that
\begin{equation}
k^{\mu} d\sigma_{\mu} =  A_{\perp} m_{\perp} \tau \cosh(\eta-y)=
A_{\perp} |
dk_{\eta} |
\end{equation}
Thus we can rewrite our expression for the field theory particle spectra
as
\begin{equation}
{d^3N \over  \pi dy\,dk_{\perp}^2} = A_{\perp} \int dk_{\eta}
f(k_{\eta}, k_{\perp},\tau)= \int f(k_{\eta}, k_{\perp},\tau) k^{\mu}
d\sigma_{\mu}
\end{equation}
where in the second integration we keep $y$ and $\tau$ fixed.
This shows that the Cooper-Frye formula for the final spectra of
secondaries,
which was derived from classical transport theory, also applies in field
theory
when one has a gaussian density matrix as in the large-N approximation.
It
also reconfirms the idea that the interpolating phase space number
density
plays
the role of a classical transport phase space density function,
as was found in our calculation of pair production from strong
electric fields \cite{kluger}

In Figures 2 and 3 we compare the boost invariant hydrodynamical result
for the
transverse momentum distribution using   critical temperatures of
$ T_c = 140, 200$
MeV to the two nonequilibrium  cases represented in figure 1.
Figure 2 pertains to the initial condition ${\dot \sigma} |_{\tau_0}=-1$
In this case there is a regime where the effective mass becomes negative
and we see  a noticable enhancement
of the low transverse momentum spectra. We have  normalized both results
to give the same total center of mass energy $E_{cm}$.
Figure 3 corresponds to the initial condition ${\dot \sigma}
|_{\tau_0}=0$.
Here we notice that there is a little enhancement at low transverse
momenta.

In conclusion, we see that a non-equilibrium phase transition
taking place during a time evolving quark-gluon or hadronic plasma
can lead to an enhancement of the low momentum transverse momentum
distribution. In particular, if a Centauro type event is not
accompanied by such an enhancement one would be suspicious of ascribing
this event to the production of disoriented chiral condensates as a
result of a rapid quench.

FIGURE 1. Proper time evolution of the $\chi$ field
for two different initial conditions
($\dot{\sigma}(1)=0$ and $\dot{\sigma}(1)=-1$)
with $f_{\pi}=92.5 MeV$.

FIGURE 2. Single particle distribution  $k_{\perp} {E dN \over d^3 k}$
vs. $k_{\perp}$ in the mean field approximation for the initial
condition $\dot{\sigma}(1)=-1$ compared to two different hydrodynamical
calculations with two different breakup temperatures:
$T_c=m_{\pi}$ and $T_c=1.4m_{\pi}$. We have normalized all the curves to
give
the same center of mass energy.

FIGURE 3. Same comparison as in fig. 2 but with the
mean field approximation for the initial
condition $\dot{\sigma}(1)=0$.


\begin{thebibliography}{9}

\bibitem{Anselm} A. Anselm, Phys. Lett. {\bf B217}, 169 (1989); A.
Anselm
and M. Ryskin, Phys. Lett. {\bf B226}, 482 (1991).

\bibitem{BjorkenIJMP} J.D. Bjorken, Int. Jour. of Mod. Phys.{\bf A7},
4189 (1992); Acta Phys. Pol. {\bf B23} 561 (1992).

\bibitem{RajaWil} K. Rajagopal and F. Wilczek, Nucl. Phys. {\bf B379},
395 (1993); S. Gavin, A. Gocksch and R.D. Pisarski, Phys. Rev. Lett.
{\bf72},2143
(1994)

\bibitem{Cosmicrays} C.M.G. Lattes, Y. Fujimoto and S. Hasegawa, Phys.
Rep.
{\bf 65}, 151 (1980).


 \bibitem{DCC} F. Cooper, Y.Kluger, E.Mottola, J.P. Paz, Phys. Rev. {\bf
D51},2377  (1995).
\bibitem{Gavin} Sean Gavin, hep-ph/9407368 and references therein.
\bibitem{cfs} F. Cooper, G. Frye and E. Schonberg, Phys. Rev. {\bf D11}
192 (1975);
J.D. Bjorken, Phys. Rev. {\bf D27}, 140 (1983)
\bibitem{c-f} F. Cooper and G. Frye Phys. Rev. {\bf D10} ,186 (1974)

\bibitem{kluger}   F. Cooper, J. M. Eisenberg, Y. Kluger, E. Mottola,
and B. Svetitsky,
Phys Rev. {\bf D 48}  190 (1993)

\end{thebibliography}
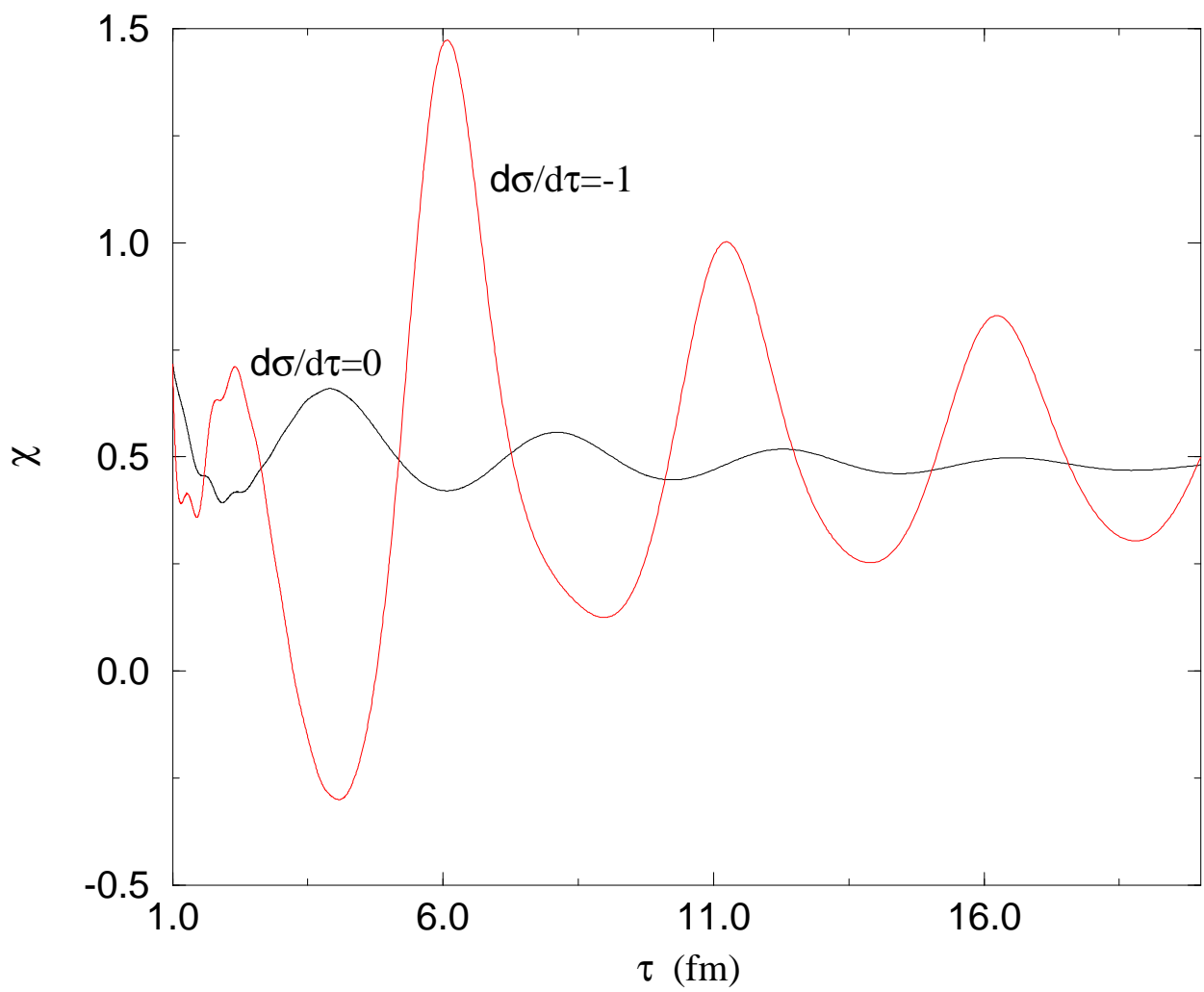
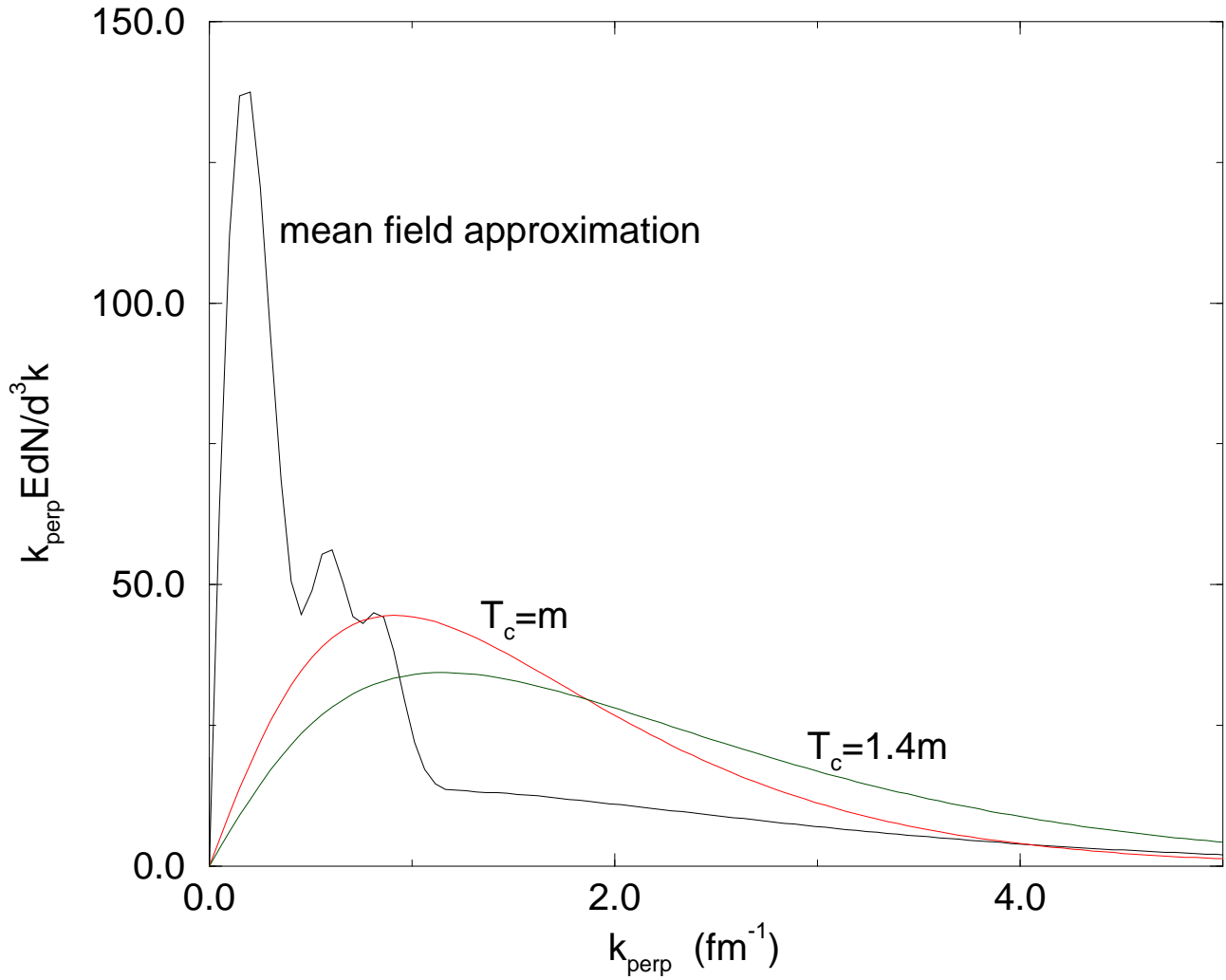
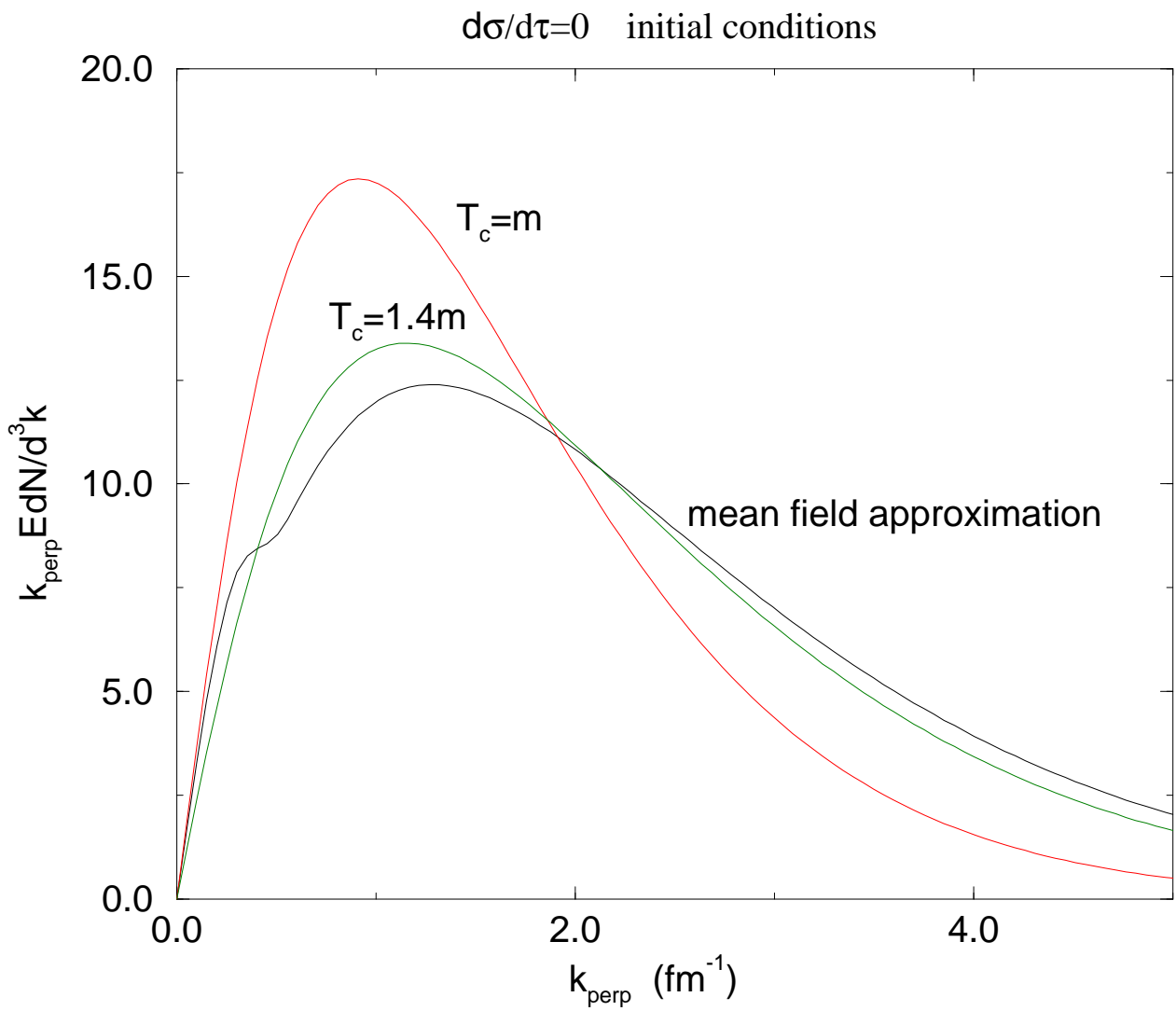
\end{document}